\documentclass[aps,prl,twocolumn,groupedaddress,nofootinbib,superscriptaddress,showpacs]{revtex4-2}

\usepackage[hidelinks]{hyperref}       
\usepackage{url}            
\usepackage{booktabs}       
\usepackage{amsfonts}       
\usepackage{amssymb}
\usepackage{mathtools}
\usepackage{nicefrac}       
\usepackage{microtype}      
\usepackage{amsmath}
\usepackage{xspace}
\usepackage[T1]{fontenc}
\usepackage{dsfont}
\usepackage{pifont}
\usepackage[export]{adjustbox}
\usepackage{relsize}
\usepackage{xcolor,colortbl}
\usepackage{hhline}
\usepackage{makecell}
\usepackage{physics}
\usepackage{bbm}
\usepackage{changes}
\usepackage{tabularx}
\usepackage{amsthm}
\usepackage[nameinlink]{cleveref}
\usepackage{changes}
\usepackage{tikz}
\usepackage{color,soul}
\usepackage{multirow}
\usepackage{braket}

\newcommand*\xbar[1]{%
  \hbox{%
    \vbox{%
      \hrule height 0.5pt 
      \kern0.5ex
      \hbox{%
        \kern-0.05em
        \ensuremath{#1}%
        \kern-0.15em
      }%
    }%
  }%
} 
  
\def\eg{\emph{e.g.}}
\def\ie{\emph{i.e.}}
\def\ii{\mathrm{i}}
  \definecolor{masoncolor}{rgb}{0.98, 0.27, 0.62}

\begin{document}
\title{Clustering-induced localization of quantum walks on networks}
\date{\today}
%
%
\author{Lucas B\"ottcher}
\email{l.boettcher@fs.de}
\affiliation{Department of Computational Science and Philosophy, Frankfurt School of Finance and Management, 60322 Frankfurt am Main, Germany}
\affiliation{Department of Medicine, University of Florida, Gainesville, FL, 32610, United States of America}
\author{Mason A. Porter}
\email{mason@math.ucla.edu}
\affiliation{Department of Mathematics, University of California, Los Angeles, CA, 90095, United States of America}
\affiliation{Department of Sociology, University of California, Los Angeles, CA, 90095, United States of America}
\affiliation{Santa Fe Institute, Santa Fe, NM, 87501, United States of America}

\date{\today}
\begin{abstract}
Quantum walks on networks are a paradigmatic model in quantum information theory. Quantum-walk algorithms have been developed for various applications, including spatial-search problems, element-distinctness problems, and node centrality analysis. Unlike their classical counterparts, the evolution of quantum walks is unitary, so they do not converge to a stationary distribution. However, for many applications, it is important to understand the long-time behavior of quantum walks and the impact of network structure on their evolution. In the present paper, we study the localization of quantum walks on networks. We demonstrate how localization emerges in highly clustered networks that we construct by recursively attaching triangles, and we derive an analytical expression for the long-time inverse participation ratio that depends on products of eigenvectors of the quantum-walk Hamiltonian. Building on the insights from this example, we then show that localization also occurs in Kleinberg navigable small-world networks and Holme--Kim power-law cluster networks. Our results illustrate that local clustering, which is a key structural feature of networks, can induce localization of quantum walks.
\end{abstract}
\maketitle
%


Quantum walks have numerous and diverse applications, including spatial-search problems~\cite{childs2004spatial,magniezsearch2011,portugal2013quantum,krovi2016quantum}, element-distinctness problems~\cite{ambainis2007quantum}, and node-centrality analysis~\cite{sanchez2012quantum,rossi2014node,izaac2017centrality,wald2020classical,boettcher2021classical,tang2021tensorflow,bottcher2024complex}. Potential applications of quantum walks extend well beyond these examples, as any problem that can be solved by a general-purpose quantum computer can also be implemented as a quantum walk on a network~\cite{childs2009universal,lovett2010universal}. Quantum walks are also effective models of various transport processes, such as energy transport in photosynthetic complexes~\cite{mohseni2008environment,rebentrost2009environment}, and provide a useful framework to examine relationships between structural and dynamical features in quantum networks~\cite{nokkala2023}. Unlike classical random walks, quantum walks evolve unitarily and thus do not converge to a stationary distribution~\cite{kempe2003quantum}. To mathematically characterize the evolution of quantum walks, one usually considers long-time means of quantities such as occupation probabilities and transition probabilities. Examining the long-time behavior of quantum walks and the influence of network structure on their evolution can help guide the integration of quantum walks into algorithms~\cite{chakraborty2020fast}. 

Dynamical processes on networks can exhibit \emph{localization}, which occurs when they are eventually confined to a small set of nodes of a network~\cite{sood2007localization,burda2009localization,goltsev2012localization,shukla2024localization,pradhan2025graph,chen2025critical}. Localization arises from the interplay between a network's structural properties and dynamical processes on it. In the present paper, we study the localization of continuous-time quantum walks (CTQWs) on networks.

Prior research on CTQWs has illustrated that adding edges uniformly at random to a ring network with nearest-neighbor connections is associated with ensemble-averaged transition probabilities that are large for an initially excited node (\ie, a single node that is initially occupied) and close to $0$ for the other nodes~\cite{mulken2007quantum}. Individual realizations of quantum walks on such networks do not exhibit localization. Averaging across different instantiations of the random edge-addition process is associated with transitions of an initial excitation of one node to any other node. These transitions tend to cancel each other out, causing excitations to remain localized on average in their initial location.
Researchers have observed localization of individual quantum walks in tree networks with Hamiltonians that incorporate disorder~\cite{keating2007localization,jackson2012quantum}. Other researchers have associated certain local connectivity patterns that arise from $0$ eigenvalues with quantum-walk localization~\cite{bueno2020null}. It is also known that quadratic perturbations can induce localization of quantum walks~\cite{candeloro2020continuous}. 

There is also a body of work on discrete-time quantum walks on networks~\cite{tregenna2003controlling,inui2004localization,inui2005one,watabe2008limit,vstefavnak2008recurrence,vstefavnak2008recurrence,schreiber2011decoherence,werner2013localization,kollar2015strongly,lyu2015localization,danaci2021disorder,sharma2022transport,duda2023quantum}. These studies include many papers on localization involving both various lattices (and a fractal network) and ``coin operators'', which act on the spin component of an underlying product state. 

To study the localization of CTQWs on networks, we first establish how localization emerges in highly clustered networks that we construct by recursively attaching triangles. We then build on this example to show
that localization also occurs in Kleinberg navigable small-world networks \cite{kleinberg00} and Holme--Kim power-law cluster networks~\cite{holme2002growing}. These localization effects have potential implications for the use of quantum walks in quantum memory, where localized walks help reduce the size of the position space that is needed to store information~\cite{chandrashekar2015localized}. Our results also illustrate how structural characteristics of networks can suppress the speed-ups of quantum-walk propagation over classical-walk propagation on networks, a factor that is relevant when employing quantum walks in quantum communication channels~\cite{keating2007localization}.


\paragraph*{Quantum walks on networks.}
\begin{figure*}
    \centering
    \includegraphics[width=\textwidth]{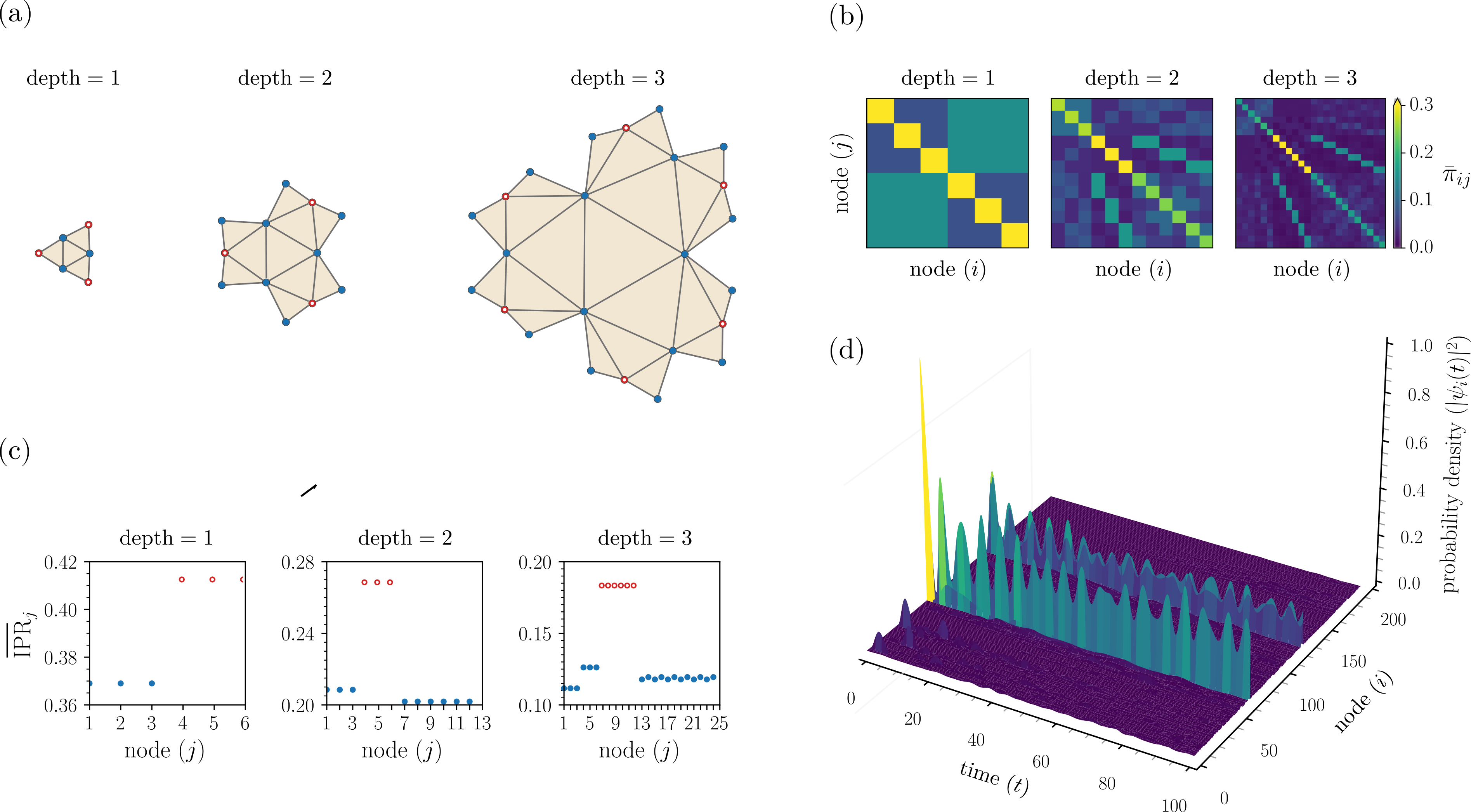}
    \caption{Localization in recursive triangle networks. (a) Recursive triangle networks of depths 1, 2, and 3. The long-time mean inverse participation ratios (IPRs) of the nodes with red rings are substantially larger than those of other nodes. (b) The long-time mean transition probabilities $\bar{\pi}_{ij}$ [see Eq.~\eqref{eq:pi_bar_ij}] for the networks in (a). (c) The long-time mean IPR $\xbar{\mathrm{IPR}}_j$ [see Eq.~\eqref{eq:IPR_bar_j}] as a function of the initially excited node $j$. The hollow red circles indicate the maximum value of $\xbar{\mathrm{IPR}}_j$. 
    (d) The probability density $|\psi_i(t)|^2 = |\braket{i|\psi(t)}|^2$ as a function of the node $i$ and time $t$ for a CTQW that starts at node 61 in a recursive triangle network of depth 6. In this example, the quantum walker predominantly alternates between two nodes.
    }
    \vspace{-1.5em}
    \label{fig:triangle_network}
\end{figure*}
We consider unweighted, undirected networks in the form of graphs $G = (V, E)$, where $V$ is a set of nodes and $E$ is a set of edges. The number of nodes is $N = |V|$, and the number of edges is $M = |E|$. We describe the edges between nodes using an adjacency matrix $A \in \{0, 1\}^{N \times N}$. The element $a_{ij}$ of the matrix $A$ is $1$ if nodes $i$ and $j$ are adjacent to each other and $0$ if they are not. Because the network is undirected, $a_{ij} = a_{ji}$. The degree of a node $i$ is $k_i = \sum_{j = 1}^N a_{ij}$. Unless we state otherwise, we do not consider self-edges (\ie, we set $a_{ii} = 0$). The mean clustering coefficient of an undirected network $G$ is $C(G) = N^{-1}\sum_{i=1}^N c_i$, where the local clustering coefficient of node $i$ is $c_i = \sum_{j,k} a_{ij}a_{jk}a_{ki}/(k_i (k_i - 1))$ for $k_i\geq 2$ and $c_i = 0$ otherwise~\cite{newman2018networks}.

The wave function $\ket{\psi} \in \mathbb{C}^N$ of a CTQW evolves according to the Schr\"odinger equation
\begin{equation}
    \frac{\mathrm{d}}{\mathrm{d}t}\ket{\psi(t)} = -\ii H \ket{\psi(t)}\,,
\end{equation}
where $\ii = \sqrt{-1}$ is the imaginary unit and the Hamiltonian $H$ is the infinitesimal generator of time translation. We assume the normalization $\braket{\psi(t)|\psi(t)} = 1$. In accordance with Refs.~\cite{faccin2013degree,wald2020classical,boettcher2021classical}, we let $H$ be the symmetric and normalized graph Laplacian matrix. That is, 
\begin{equation}
    H = D^{-1/2} L D^{-1/2}\,,
    \label{eq:symm_graph_lap}
\end{equation}
where $D = \sum_{j = 1}^N k_j\ket{j}\bra{j}$ is the degree matrix and $L = D-A$ is the combinatorial graph Laplacian matrix. The quantity $\ket{j}\in\mathbb{C}^N$ is an orthonormal basis vector that satisfies $\braket{i|j} = \delta_{ij}$, where $\delta_{ij}$ denotes the Kronecker delta, which is $1$ if $\ket{i} = \ket{j}$ and is $0$ otherwise. For the choice \eqref{eq:symm_graph_lap} of $H$, if a system is in the ground state, then the probability of finding a quantum walker on a given node is the same as in a classical random walk with the generator $LD^{-1}$~\cite{faccin2013degree}.
%


\paragraph*{Localization measures.}
The probability that a CTQW with Hamiltonian $H$ transitions from node $j$ at time $0$ to node $i$ at time $t$ is
\begin{align}
\begin{split}
    \pi_{ij}(t) &= |\braket{i|e^{-\ii H t}|j}|^2\\
    		&= \sum_{m=1}^N A^{ij}_m+\sum_{m<n} B^{ij}_{mn} \cos((\lambda_m - \lambda_n) t)\,,
\end{split}
\label{eq:pi_ij}
\end{align}
where $A^{ij}_m\coloneqq |\braket{i|e_m}\braket{e_m|j}|^2$ and $B^{ij}_{mn}\coloneqq 2\braket{i|e_m}\braket{e_m|j} \braket{j|e_n}\braket{e_n|i}$. The quantities $\lambda_m$ and $e_m$, respectively, are the eigenvalues and corresponding eigenvectors of $H$. That is, $H e_m = \lambda_m e_m$. The long-time mean of the transition probability $\pi_{ij}(t)$ is
\begin{equation}
    \bar{\pi}_{ij} = \lim_{T\rightarrow\infty}\frac{1}{T}\int_0^T \pi_{ij}(t)\,\mathrm{d}t\,.
\label{eq:pi_bar_ij}
\end{equation}
Inserting Eq.~\eqref{eq:pi_ij} into Eq.~\eqref{eq:pi_bar_ij} yields
\begin{align}
    \bar{\pi}_{ij} &= \sum_{m=1}^N A^{ij}_m + \sum_{\substack{m < n \,, \\ \lambda_m = \lambda_n}} B^{ij}_{mn}\,.
\end{align}

To quantify the amount of localization of a CTQW, we calculate the inverse participation ratio (IPR)~\cite{thouless1974electrons,wegner1980inverse} 
\begin{equation}
    \mathrm{IPR}_j(t) = \sum_{i = 1}^N |\braket{i|e^{-\ii H t}|j}|^4 = \sum_{i = 1}^N \pi_{ij}^2(t)
\end{equation}
that is associated with the initial state $\ket{j}$. To interpret the IPR, consider a wave function in which $\ell$ elements have magnitude $1/\sqrt{\ell}$ and $N - \ell$ elements have magnitude $0$. This wave function has an IPR of $\ell(1/\sqrt{\ell})^4 = 1/\ell$. For a fully localized state, in which a CTQW is localized at a single node (\ie, $\ell = 1$), the IPR attains its maximum value of $1$. For a fully delocalized state (\ie, $\ell = N$), the IPR attains its minimum value of $1/N$.

The long-time mean of $\mathrm{IPR}_j(t)$ is
%
\begin{align}
\begin{split}
    \xbar{\mathrm{IPR}}_j &= \lim_{T\rightarrow\infty}\frac{1}{T}\int_0^T \mathrm{IPR}_j(t)\,\mathrm{d}t\\
    &= \sum_{i = 1}^N\Big[\Big(\sum_{m = 1}^N A^{ij}_m\Big)^2 + 2\sum_{m = 1}^N A^{ij}_m \sum_{\substack{m < n\,, \\ \lambda_m = \lambda_n}} B^{ij}_{mn}\\
    &\hspace{3em} + \sum_{\substack{m < n\,, \,r < s \,, \\ \lambda_m - \lambda_n = \lambda_r - \lambda_s}} C_{mnrs}^{ij} + \sum_{\substack{m < n\,, \, r < s \,, \\ \lambda_m - \lambda_n = \lambda_s - \lambda_r}} C_{mnrs}^{ij} \Big]\,,
\label{eq:IPR_bar_j}
\end{split}
\end{align}
%
where $C_{mnrs}^{ij} = B^{ij}_{mn}B^{ij}_{rs}/2$. The second term of $\xbar{\mathrm{IPR}}_j$ measures localization due to degenerate eigenvalues, and the third and fourth terms quantify localization that is associated with eigenvalue quartets satisfying $\lambda_m - \lambda_n = \lambda_r - \lambda_s$ and $ \lambda_m - \lambda_n = \lambda_s - \lambda_r$, respectively.


\paragraph*{Localization in recursive triangle networks and related networks.}
As a starting point, we examine recursive triangle networks [see Fig.~\ref{fig:triangle_network}(a)]. These networks and related networks have also been studied in the framework of ``network geometry with flavor'' (NGF)~\cite{bianconi2016network,bianconi2017emergent,bianconi2020spectral} and in the context of hyperbolic lattices~\cite{PhysRevLett.133.061603,chen2024anderson}. For a given depth $d$, recursive triangle networks have $N = 3\times 2^d$ nodes and $M = 3\times (2^{d + 1} - 1)$ edges. In Fig.~\ref{fig:triangle_network}(a), we show such networks with depths of $d=1$, $d = 2$, and $d = 3$. The mean clustering coefficients of these networks are 0.75, 0.71, and 0.70, respectively. For larger values of $d$, the mean clustering coefficient approaches a value of approximately 0.69.

For the recursive triangle networks with depths $d = 1$, $d = 2$, and $d = 3$, we show the long-time mean transition-probability elements $\bar{\pi}_{ij}$ for all $i,j\in\{1,\ldots,N\}$ in Fig.~\ref{fig:triangle_network}(b). For depth $d = 1$, one can obtain analytical expressions for the eigenvalues and eigenvectors of the Hamiltonian \eqref{eq:symm_graph_lap}. This enables one to write down the long-time mean transition-probability matrix
\begin{equation}
\bar{\boldsymbol{\pi}}^{(d=1)} = \frac{1}{27} \begin{pmatrix}
	11 & 2 & 2 & 4 & 4 & 4 \\
	2 & 11 & 2 & 4 & 4 & 4 \\
	2 & 2 & 11 & 4 & 4 & 4 \\
	4 & 4 & 4 & 11 & 2 & 2 \\
	4 & 4 & 4 & 2 & 11 & 2 \\
	4 & 4 & 4 & 2 & 2 & 11 \\
\end{pmatrix}\,.
\end{equation}
All diagonal elements $\bar{\pi}_{ii}^{(d = 1)}$ have the same value and are much larger than the off-diagonal elements $\bar{\pi}_{ij}^{(d = 1)}$ (with $i\neq j$). This indicates that a CTQW that starts at node $j$ has a higher probability of revisiting node $j$ over a long time horizon than of occupying any other node. For recursive triangle networks with depths 2 and 3, over a long time horizon, CTQWs that start from nodes 4--6 and 7--12, respectively, have higher probabilities of revisiting these nodes than of occupying any other node [see Fig.~\ref{fig:triangle_network}(b)].

In Figs.~\ref{fig:triangle_network}(a,c), we highlight nodes with hollow red circles when their long-time mean IPRs [see Eq.~\eqref{eq:IPR_bar_j}] are much larger than those of other nodes. For the recursive triangle network with depth $d = 1$, nodes 1--3 have a long-time mean IPR of $269/729 \approx 0.37$ and nodes 4--6 have a long-time mean IPR of $301/729 \approx 0.41$. For the recursive triangle networks with depths 2 and 3, the nodes with the largest long-time mean IPRs are the nodes with the largest diagonal elements of $\bar{\pi}_{ij}$ in Fig.~\ref{fig:triangle_network}(b). 

\begin{figure}
    \centering
    \includegraphics{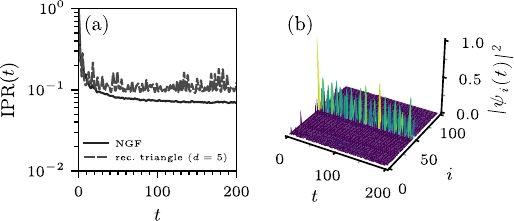}
    \caption{Localization on networks that we construct using the network geometry with flavor (NGF) model. (a) Evolution of the IPR for both NGF networks (with $s = -1$) and a recursive triangle network with depth $d = 5$. For the NGF model, the curve is the mean of the results for 1000 NGF networks, which each have 100 nodes. The CTQW starts at node 51 in the NGF networks. The recursive triangle network has 96 nodes, and the CTQW starts at node 36, where localization is strongest. (b) The probability density $|\psi_i(t)|^2 = |\braket{i|\psi(t)}|^2$ as a function of the node $i$ and time $t$ for a CTQW that starts at node 51 in one 100-node NGF network (with $s = -1$).
    }
    \vspace{-2em}
    \label{fig:ngf}
\end{figure}

The Hamiltonian $H$ of the recursive triangle network with depth $d = 1$ has eigenvalues 0, 3/4, 3/4, 3/2, 3/2, and 3/2. {The degeneracies of the eigenvalues 3/4 and 3/2 are associated with contributions of the sums over $B_{mn}^{ij}$ with $\lambda_m = \lambda_n$ [\ie, the second term in Eq.~\eqref{eq:IPR_bar_j}]. The sums over $C_{mnrs}^{ij}$ [\ie, the third and fourth terms in Eq.~\eqref{eq:IPR_bar_j}] also contribute to the overall long-time mean IPR, as several eigenvalue quartets satisfy 
$\lambda_m - \lambda_n = \lambda_r - \lambda_s$ and $\lambda_m - \lambda_n = \lambda_s - \lambda_r$. We observe similar characteristics in recursive triangle networks with larger depths. For example, the eigenvalue distributions of those networks also have a degeneracy for the eigenvalue 3/2 that is associated with contributions from the second term in Eq.~\eqref{eq:IPR_bar_j}. These results illustrate how localization is associated with both eigenvalue degeneracies and quartet spacings. It may also be useful to examine how localization relates to the spectral dimension, which has been used to link eigenvalue statistics with the dynamics of dynamical processes on networks~\cite{millan2018complex, millan2019synchronization}.

\begin{figure}
    \centering
    \includegraphics{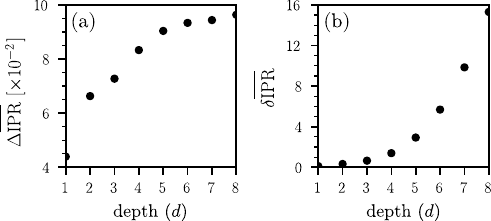}
    \caption{The absolute and relative gaps between the maximum and minimum long-time mean IPRs for recursive triangle networks. (a) The absolute gap $\Delta \xbar{\mathrm{IPR}}$ between the maximum and minimum long-time mean IPRs [see Eq.~\eqref{eq:Delta_IPR}] as a function of the network depth $d$.
    (b) The relative gap $\delta \xbar{\mathrm{IPR}}$ between the maximum and minimum long-time mean IPRs [see Eq.~\eqref{eq:delta_IPR}] as a function of the network depth $d$.
    }
    \vspace{-1.5em}
    \label{fig:IPR_gaps}
\end{figure}

We also examine the absolute gap
\begin{equation}
    \Delta \xbar{\mathrm{IPR}}=\max_j (\xbar{\mathrm{IPR}}_j) - \min_j (\xbar{\mathrm{IPR}}_j)
    \label{eq:Delta_IPR}
\end{equation}
and relative gap
\begin{equation}
    \delta \xbar{\mathrm{IPR}}=\frac{
    \max_j (\xbar{\mathrm{IPR}}_j) - \min_j (\xbar{\mathrm{IPR}}_j)}{\min_j (\xbar{\mathrm{IPR}}_j)}\,,
    \label{eq:delta_IPR}
\end{equation}
between the maximum and minimum long-time mean IPRs for recursive triangle networks with different depths. Consistent with the trend that we observed in Fig.~\ref{fig:triangle_network}(c), the values of both $\Delta \xbar{\mathrm{IPR}}$ and $\delta \xbar{\mathrm{IPR}}$ increase with the depth $d$ [see Fig.~\ref{fig:IPR_gaps}]. For $d = 1$, the absolute gap is $\Delta \xbar{\mathrm{IPR}} \approx 0.04$ and the relative gap is $\delta \xbar{\mathrm{IPR}} \approx 0.11$. By contrast, for $d = 8$, the absolute gap is $\Delta \xbar{\mathrm{IPR}} \approx 0.10$ and the relative gap is $\delta \xbar{\mathrm{IPR}} \approx 15.30$. The dependence of $\Delta \xbar{\mathrm{IPR}}$ and $\delta \xbar{\mathrm{IPR}}$ on the depth $d$ in Fig.~\ref{fig:IPR_gaps} suggests that the absolute and relative IPR gaps both approach a finite value {as $d \rightarrow \infty$} but that $\min_j (\xbar{\mathrm{IPR}}_j)$ approaches $0$ {as $d \rightarrow \infty$}. For larger depths, some node excitations become highly localized, but others become increasingly delocalized.

As an example of a CTQW with substantial localization, we show the CTQW evolution on a recursive triangle network with depth 6 in Fig.~\ref{fig:triangle_network}(d). In this simulation, the CTQW starts at node 61 (of $3 \times 2^6 = 192$ nodes) and has $\xbar{\mathrm{IPR}}_{61} \approx 0.11$. A closer examination of the probability density oscillations reveals that the quantum walker predominantly alternates between two nodes.

To examine the impact of randomness in network generation on recursive triangle networks, which are deterministic, we consider two-dimensional (2D) NGF networks with a flavor value of 
$s = -1$~\cite{bianconi2015complex,bianconi2016network,bianconi2017emergent}. 
In these NGF networks, a flavor value of $s = -1$ constrains each edge to be  adjacent to at most two triangles. (NGF networks with other flavor values do not have this constraint.) Rather than attaching new triangles deterministically, the NGF model creates networks through a stochastic process. In our simulations, we set the model's inverse-temperature parameter to $0$, which yields the Eden model~\cite{bottcher2021computational} on a 2D simplicial complex~\cite{bianconi2015complex}. In Fig.~\ref{fig:ngf}(a), we compare the evolution of the IPR for NGF networks with $s = -1$ and a recursive triangle network with depth $d = 5$. We we average the IPR over 1000 NGF networks. In the NGF networks (which have 100 nodes), the CTQW starts at node 51; in the recursive triangle network (which has 96 nodes), the CTQW starts at node 36, where localization is strongest. The mean clustering coefficient of these NGF networks is about 0.65, and it is about 0.69 for the recursive triangle network. The mean IPR of the NGF networks are similar in magnitude to the IPR of the recursive triangle network. To further illustrate localization dynamics, Fig.~\ref{fig:ngf}(b) shows the CTQW evolution on a single NGF network. The randomness in the NGF generation process reduces the number of degenerate eigenvalues. For example, the 100-node NGF network in Fig.~\ref{fig:ngf}(b) has one degenerate eigenvalue, whereas the 96-node recursive triangle network has 22 degenerate eigenvalues.
Consequently, the associated terms in Eq.~\eqref{eq:IPR_bar_j} contribute less to the long-time mean IPR in NGF networks than in recursive triangle networks.}

\paragraph*{Localization in other networks.}
\begin{figure}
    \centering
    \includegraphics{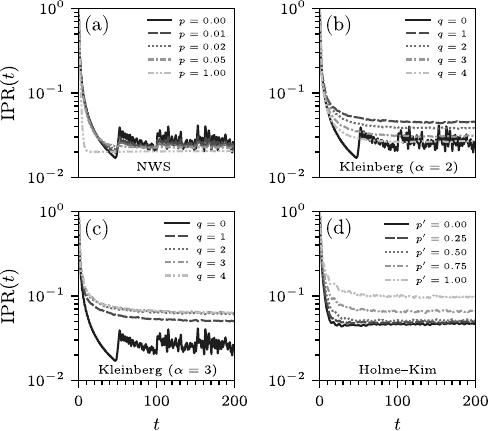}
    \caption{Evolution of the IPR for several types of networks. (a) A Newman--Watts--Strogatz (NWS) network in which each node is adjacent to its two nearest neighbors in a ring. 
    For each edge, the probabilities of adding a new edge are $p \in \{0, 0.01, 0.02, 0.05, 1\}$. (b, c) A Kleinberg navigable small-world network with clustering exponent $\alpha$ and $q \in \{0,1,2,3,4\}$ additional connections for each node. In (b), $\alpha = 2$; in (c), $\alpha = 3$. (d) A Holme--Kim (HK) power-law cluster network with probability $p' \in \{0,0.25,0.5,0.75,1\}$ of adding a triangle after adding a random edge. We construct this network by starting with an empty dyad (\ie, two isolated nodes) and iteratively adding new nodes until the network has 100 nodes. Each new node connects to two existing nodes using linear preferential attachment. All networks have $N = 100$ nodes, and all curves are means of results for 1000 networks. The CTQWs start at node 51.
    }
    \vspace{-2em}
    \label{fig:ipr_plot}
\end{figure}

To further study localization, we consider three additional types of networks: (i) Newman--Watts--Strogatz (NWS) small-world networks~\cite{watts1998collective,newman1999renormalization}, (ii) Kleinberg navigable small-world networks~\cite{kleinberg00} with a ring structure, and (iii) Holme--Kim (HK) power-law cluster networks~\cite{holme2002growing}. In the NWS networks, each node is adjacent to its two nearest neighbors in a ring. 
For each existing edge between nodes $i$ and $j$, with probability $p$, we add a new edge between node $i$ and another node that we select uniformly at random. In a Kleinberg network, edges that one adds to the backbone ring are more likely to connect nodes that are closer to each other (with respect to distance in the ring) than those that are farther apart from each other. When adding an edge, the probability of connecting two nodes $i$ and $j$ is proportional to $d(i,j)^{-\alpha}$, where $\alpha \geq 0$ is the clustering exponent and $d(i,j)$ is the geodesic distance between nodes $i$ and $j$. (Observe that self-edges are possible.) The HK network is a generalization of the standard Barab\'asi--Albert preferential-attachment model \cite{newman2018networks} that also adds triangles. One starts with an empty dyad (\ie, two isolated nodes). In each preferential-attachment step, one adds a new node and connects it to two existing nodes, which one chooses with probabilities that are proportional to their degrees. With probability $p'$, for each new node and incident edge from the preferential-attachment step, one adds another edge and forms a triangle by connecting the new node to a neighbor of the previously linked node.

\begin{figure}
    \centering
    \includegraphics{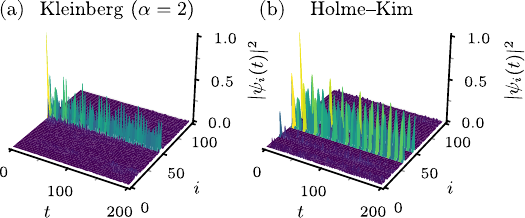}
    \caption{
    The probability density $|\psi_i(t)|^2 = |\braket{i|\psi(t)}|^2$ as a function of the node $i$ and time $t$ for a CTQW that starts at node 51 in (a) a 100-node Kleinberg network (with $\alpha = 2$ and $q = 1$) and (b) a 100-node Holme--Kim network (with $p' = 1$).
    }
    \vspace{-2em}
    \label{fig:localization_nets}
\end{figure}

In Fig.~\ref{fig:ipr_plot}, we show the IPR as a function of time for NWS, Kleinberg, and HK networks. We consider different model parameters and calculate means of results for 1000 networks. For the NWS networks, we consider the probabilities $p \in \{0, 0.01, 0.02, 0.05, 1\}$ and observe that the IPR approaches values between $0.02$ and $0.03$ [see Fig.~\ref{fig:ipr_plot}(a)], corresponding to a delocalized wave function. The means of the mean clustering coefficient of the NWS networks are $0$ for $p = 0$ and about $0.03$ for $p = 1$. When $p = 0$, the CTQW travels around the ring and interferes with itself. For the Kleinberg networks, we consider the clustering exponents $\alpha = 2$ [see Fig.~\ref{fig:ipr_plot}(b)] and $\alpha = 3$ [see Fig.~\ref{fig:ipr_plot}(c)] and we add $q$ connections per node to the underlying ring. When we do not not add any edges (\ie, $q = 0$), the observed evolution of the IPR resembles that in Fig.~\ref{fig:ipr_plot}(a) for $p = 0$ and has a mean clustering coefficient of $0$. For $\alpha = 2$ and $q = 1$, the IPR approaches values of about $0.05$, hinting at localization. Visually inspecting the wave-function evolution confirms that the CTQW has a noticeable amplitude for only a few nodes in some of the networks [see Fig.~\ref{fig:localization_nets}(a)]. For the Kleinberg networks with $\alpha = 2$ and $q \geq 1$, the IPR approaches a smaller value as we increase $q$ [see Fig.~\ref{fig:ipr_plot}(b)]. For $\alpha = 2$, it is likely that the edge-addition process introduces some long-range edges, causing the CTQW to propagate through the whole network rather than being confined to a local region. (This is somewhat reminiscent of the appearance of new infection clusters via long-range edges in spreading process on networks~\cite{taylor2015}.) For $\alpha = 3$, the added edges are more likely to connect nearby nodes than to connect distant nodes. In this case, the IPR approaches values between $0.05$ and $0.06$ as we increase $q$ [see Fig.~\ref{fig:ipr_plot}(c)].  We believe that this is due to the existence of many local clusters in the networks. For both $\alpha = 2$ and $\alpha = 3$, the mean of the mean clustering coefficients of the Kleinberg networks approaches a value between $0.28$ and $0.29$ for $q = 4$. In the HK networks, for $p'$ close to 1, we observe IPR values of up to $0.1$ [see Fig.~\ref{fig:ipr_plot}(d)]. This indicates that we observe localization, which is further evident in the evolution of individual wave functions [see Fig.~\ref{fig:localization_nets}(b)]. The mean of the mean clustering coefficients of the HK networks is about $0.13$ for $p' = 0$ and about $0.74$ for $p' = 1$.
%
%
%

\paragraph*{Conclusions and discussion.}
We examined how local clustering can induce localization in quantum walks on networks, and we thereby obtained insights into the influence of network structure on their long-time behavior.
A key result is that local clustering can inhibit the propagation of quantum walks, which is relevant to consider when employing quantum walks in quantum communication channels~\cite{keating2007localization}. 
We also demonstrated that the inverse participation ratio (IPR) is a useful indicator of localization. However, identifying a universal IPR threshold that clearly distinguishes between localized and delocalized states is not straightforward, so it is also useful to visually inspect wave-function evolution.
 
Our findings motivate experimental studies of quantum-walk localization in clustered networks using, \eg, photonic implementations of quantum  walks~\cite{tang2018experimental}. Such experiments can probe the role of local clustering in directing quantum-transport processes. In this context, it is worthwhile to study localization effects for the potential applications of quantum walks in quantum memory, as localized quantum walks require less space than non-localized quantum walks to store and retrieve information~\cite{chandrashekar2015localized}.


%

\paragraph*{Acknowledgements.} 
LB acknowledges financial support from hessian.AI.


%
\bibliography{refs-v11.bib}
%
%


%
\end{document}